\documentstyle[twoside,fleqn,espcrc2]{article}

\newcommand{\beq}{\begin{equation}}
\newcommand{\eeq}{\end{equation}}
\newcommand{\bea}{\begin{eqnarray}}
\newcommand{\eea}{\end{eqnarray}}
\newcommand{\st}{\theta}

\newcommand{\AmS}{{\protect\the\textfont2
  A\kern-.1667em\lower.5ex\hbox{M}\kern-.125emS}}

\title{{Finite-size scaling on the Ising coexistence line}%
 \thanks{Presented by A.~Irb\"ack}}
\author{Sourendu Gupta\address{HLRZ, c/o KFA J\"ulich \\
        D-5170 J\"ulich, Germany}
        and
        A.~Irb\"ack\address{Theory Division, CERN \\
        CH-1211 Geneva 23, Switzerland}}

\begin{document}

\begin{abstract}
We study the finite-size scaling of moments of the
magnetization in the low-temperature phase of the
two-dimensional Ising model.
\end{abstract}

\maketitle

\section{INTRODUCTION}

Recently, finite-size scaling (FSS) at first-order phase transitions has
attracted considerable interest.
Its practical importance in connection with Monte Carlo studies
has become increasingly clear, for instance through the
controversy about the order of the phase transition in
quenched QCD. At the same time, rigorous developments
have been accomplished
by Borgs, Koteck\'y and Miracle-Sol\'e \cite{bk,bkm}.
These results were obtained for a class of models that
includes the Ising
model at large $\beta$ and $q$-state
Potts models with large $q$, and were
proposed to be more general. The FSS model suggested
by the results is quite powerful.
It provides, for instance, an estimator of
the transition coupling that should have exponentially
small finite-size corrections.
Numerical studies of Potts models support this
prediction \cite{bj,blm}, but have also found less convincing
agreement for some other quantities \cite{blm,lk}. To find out more
about the applicability of the FSS model
we have performed a numerical study
of the two-dimensional Ising model \cite{us}.

For this model, the results of ref.~\cite{bk} show
that the partition function in a large
periodic box $L^d$, at large $\beta$,
is well approximated by
\beq
Z(h,L)\,\approx\, e^{L^df_+(h)}\,+\,e^{L^df_-(h)}\,,
\label{Zper}\eeq
where $h=\beta H$, $H$ being the external magnetic field.
The functions $f_{\pm}(h)$ are related
to the free-energy density $f(h)=\lim_{L\to\infty}L^{-d}\ln Z(h,L)$
by $f_{+(-)}(h)>f_{-(+)}(h)=f(h)$ if $h>0(<0)$, and $f_+(0)=f_-(0)$.
Each term in eq.~(\ref{Zper}) represents
fluctuations about one of the two
coexisting phases, whereas the omitted remainder accounts for
mixed configurations and is exponentially suppressed in $L$.
This decomposition is the starting point
for the finite-size analysis of moments of the order parameter.
One proceeds by expanding the
functions $f_{\pm}(h)$ about $h=0$ and obtains, at fixed $hL^d$,
expansions of the moments in powers of $L^{-d}$.
The first-order finite-size corrections, to
combinations of moments of up to fourth order,
have been investigated numerically for the $q$-state Potts model with
$q=7,8$ and 10 in refs.~\cite{blm,lk}.
As already alluded to, the
agreement was not perfect, indicating that
the asymptotic $L$ regime had not been reached.
First-order corrections have also been analysed numerically for
the Ising model \cite{bl}, within the framework
of the double-Gaussian model \cite{bl,clb}. In this earlier
phenomenological FSS model, the two peaks in
the probability distribution for the order parameter were
approximated by Gaussians. This corresponds
to a quadratic approximation for $f_\pm$ in the model discussed
here, as can be seen by inverse Laplace transformation. As a result, the
double-Gaussian model produces
the same first-order expressions, except for
problems with the relative normalization of the peaks.
For the Ising model the normalization problem is absent thanks to
the global spin-flip symmetry.

We shall extend these considerations in two directions:
we shall study the size-dependence beyond the first-order level
and we shall include in the analysis moments of higher order.

\section{HIGHER ORDERS}

We consider the behaviour of the moments at the phase transition.
We calculate FSS expressions from eq.~(\ref{Zper}), assuming that
the contribution from the omitted tunnelling term is exponentially
suppressed. For definiteness, we restrict ourselves to
the Ising case.
The power-like volume dependence arises from fluctuations about the
two stable phases and is obtained by expanding the
functions $f_\pm$ about $h=0$. We write
\beq
f_+(h)=\sum_{k=0}^\infty {a_k\over k!}h^k\,,
\label{as}\eeq
and, by symmetry, the corresponding coefficients for $f_-$ then are
$(-1)^ka_k$. The coefficient $a_1=m_0$ is the spontaneous
magnetization and $\beta a_2=\chi$ the susceptibility.
In order to obtain the moment $\langle m^k\rangle$ at $h=0$ it is
sufficient to consider terms up to $k$th order in the expansions
of $f_\pm$. Inserting these expansions into eq.~(\ref{Zper}), it is
easily found, for even $k(>0)$, that
\begin{displaymath}
\langle m^k\rangle\,=\,
a_1^k\,+\,
L^{-d}\st_{k2}{k\choose 2}
a_2 a_1^{k-2}\,+\,L^{-2d}\times
\end{displaymath}\beq
\Biggl[
\st_{k3}{k\choose 3}a_3a_1^{k-3}\,+\,
\st_{k4}3{k\choose 4}
a_2^2 a_1^{k-4}\Biggr]\,+\,\ldots \,,
\label{mom}\eeq
where $\st_{ij}=1$ if $i\ge j$ and 0 otherwise.
The coefficient in front of $L^{-jd}$
is a function of $a_1,\ldots,a_{j+1}$, and it is zero
if $j\ge k$.
Odd moments, of course, vanish becuase of the spin-flip symmetry.

For moments of moderate order $k$ it is straightforward
to extend eq.~(\ref{mom}) so as to include all the power-like corrections.
The validity of the resulting expressions will however depend on $k$.
In general we expect that the (approximate) validity gets
restricted to larger $L$ with increasing $k$.
Note that higher coefficients in eq.~(\ref{mom}) can be large
not only because of the combinatorial factors, but also
because they involve higher $a_k$. The series
eq.~(\ref{as}) is asymptotic only \cite{bakerkim}.

Having obtained the prediction, we next discuss how to
apply and test it. For the two-dimensional Ising model,
the spontaneous magnetization $a_1$ is exactly known, and
fairly precise estimates of higher $a_k$'s exist in the
literature. Thus, one has a parameter-free prediction
for low-order moments. In general, the $a_k$'s are, however, unknown.
We may then try to use fits of numerical data to
eq.~(\ref{mom}) to extract them.
By checking for consistency values obtained from
different moments, a test of the FSS model is obtained.
In practice, such a program is not easy to follow, and we are
therefore led to look for other possibilities. An obvious one
is to try to use single-phase cumulants. We define the
measurement of the single-phase cumulant by introducing
a cut between the two peaks in the probability distribution
for the order parameter. In the remaining part, say $m>m^*$,
where $|m^*|<m_0$, we expect contributions from the disfavoured
phase, as defined by the decomposition of eq.~(\ref{Zper}),
as well as from mixed phases, to be exponentially
suppressed in $L$. Up to exponentially small errors, we
therefore expect
\beq
a_k\,=\,L^{(k-1)d}\langle m^k\rangle^*_c\,,
\label{cum}\eeq
where the ${}^*$ indicates expectation value over $m>m^*$.
In our calculations we take $m^*=0$, but in principle one could
choose any $|m^*|<m_0$. Note also that the applicability of
eq.~(\ref{cum}) depends on $k$. The measurement of
$\langle m^k\rangle^*_c$ rapidly
becomes more difficult with increasing order $k$, since it
is $O(L^{-(k-1)d})$ and is the result of
cancellations between $O(1)$ numbers.

\section{NUMERICAL RESULTS}

We now would like to test the finite-size expression for
moments, eq.~(\ref{mom}), and also the feasibility of a numerical
determination of the coefficients $a_k$ through eq.~(\ref{cum}).
To this end we have performed numerical
simulations of the two-dimensional Ising model.
These were done at $u/u_c=0.9$, where
$u=\exp(-4\beta)$ and $u_c=3-2\sqrt{2}$. At this temperature,
the exactly known values of the correlation length
and the spontaneous magnetisation are $\xi\approx 4.8$ and
$m_0\approx 0.84$, respectively.
In addition, several
of the coefficients $a_k$ are known to high
precision from power-series expansions \cite{bakerkim}.

Using the Swendsen-Wang cluster algorithm, we
simulated $L^2$ lattices with $16\le L\le 60$. For each size,
4 million iterations were done. Measurements were separated
by 10 iterations and errors were estimated through a
jackknife procedure.

The first three (normalized) single-phase cumulants
$c_k=L^{(k-1)d}\langle m^k\rangle^*_c$ are shown in
fig.~\ref{fig:cum}. They were obtained by
folding the probability distribution
onto $m>0$.
{}From the above discussion, we expect that
$c_k$ approaches $a_k$ at an exponential rate for
large $L$. While our data at large $L$ are consistent with
such a behaviour, it is clear that the smaller $L$
studied do not belong to the asymptotic region. Unfortunately,
the data points left are insufficient for a study of
the form of the asymptotic $L$ dependence. On the other
hand, the close agreement found at large $L$
with the values in ref.~\cite{bakerkim}
shows that good estimates of the infinite volume limit can
be obtained from the range of $L$ studied. For instance, we
note that the first and second cumulants are only
a few per cent off their infinite volume values if $L/\xi> 5$.
It would therefore be possible to determine the $L^{-d}$ term
of the moments to a similar accuracy.

\begin{figure}[p]
{\rule[-21mm]{0mm}{175mm}}
\caption{Normalized single-phase cumulants $c_k$
against $L^{-d}$. The horizontal lines are the
coefficients $a_k$, as given by ref.~[9].}
\label{fig:cum}
\end{figure}
\begin{figure}[p]
{\rule[-21mm]{0mm}{175mm}}
\caption{Moments $\langle m^k\rangle$ against $L^{-d}$.
The curves are explained in the text.}
\label{fig:mom}
\end{figure}

In fig.~\ref{fig:mom} we compare
our results for the first three even moments
with the prediction of eq.~(\ref{mom}), using
values of the $a_k$'s from ref.~\cite{bakerkim}.
For each moment all non-vanishing terms in the expansion
have been included. The curve labelled $j$ is the result
obtained by including the terms of order up to $L^{-jd}$.
At large $L$, the prediction is in very good agreement with the data,
and the agreement is improved by adding more terms.
The extent of this asymptotic $L$ region already shows,
for these three moments, a dependence on the order of the moment,
as discussed above. Larger $L$ is required for the sixth moment.
For the second and fourth moment the difference between
measured and predicted finite-size correction is less than 10\%
down to the smallest size studied, $L/\xi\approx 3.3$.
The largest discrepancy for the second moment,
at $L/\xi\approx 3.3$, is 6\% or
$3.4\sigma$.

\section{SUMMARY}

We have studied the finite-size dependence of moments
of the magnetization in the two-dimensional Ising model.
The FSS model by Borgs, Koteck\'y and
Miracle-Sol\'e gives a very good description of the large-$L$ results.
Data for
the first three even moments showed, in fact, no significant deviation
for $L/\xi>6$, after taking higher-order corrections into account.
The description remains approximately valid
at the smaller $L/\xi> 3.3$
studied for the second and fourth moment, but not for the sixth.
Existing analytical
results allowed us to perform these tests without using any free
parameters. In general the parameters of the FSS model
are unknown and we therefore tried to measure
them numerically through single-phase cumulants.
We found that the
first-order correction to the moments could be obtained
to an accuracy of a few per cent from lattices with $L/\xi>5$.

\end{document}